# Prediction and model-assisted estimation of diameter distributions using Norwegian national forest inventory and airborne laser scanning data

Janne Räty*[1], Rasmus Astrup[1] and Johannes Breidenbach*[1]

[1]Norwegian Institute of Bioeconomy Research (NIBIO), Division of Forest and Forest Resources, National Forest Inventory, Høgskoleveien 8, 1433 Ås, Norway

*Correspondence: Janne[dot]Raty[at]nibio.no, Johannes[dot]Breidenbach[at]nibio.no

## ABSTRACT

Diameter at breast height (DBH) distributions offer valuable information for operational and strategic forest management decisions. We predicted DBH distributions using Norwegian national forest inventory and airborne laser scanning data and compared the predictive performances of linear mixed- effects (PPM), generalized linear-mixed (GLM) and k nearest neighbor (NN) models. While GLM resulted in smaller prediction errors than PPM, both were clearly outperformed by NN. We therefore studied the ability of the NN model to improve the precision of stem frequency estimates by DBH classes in the 8.7 Mha study area using a model-assisted (MA) estimator suitable for systematic sampling. MA estimates yielded greater than or approximately equal efficiencies as direct estimates using field data only. The relative efficiencies (REs) associated with the MA estimates ranged between 0.95–1.47 and 0.96–1.67 for 2 and 6 cm DBH class widths, respectively, when dominant tree species were assumed to be known. The use of a predicted tree species map, instead of the observed information, decreased the REs by up to 10%.

**Keywords**: generalized linear models, linear mixed-effects models, most similar neighbor approach, number of stems, systematic sampling

## 1 INTRODUCTION

Forest inventories provide essential information for the sustainable management of forest resources at different spatial levels. A key forest attribute is the diameter at breast height (DBH) which is correlated with many other tree attributes, such as timber volume, biomass and timber assortments. Therefore, the distribution of DBHs (henceforth DBH distribution) is vital in the assessment of timber-related attributes. DBH distributions are also indicative of forest structural characteristics which may be relevant for biodiversity assessments (Valbuena et al. 2013). In the planning of forest management operations, the main purpose of DBH distributions is to characterize forest stands at the tree-level. Tree-level information is required, for example, in growth simulations when tree-level models are applied (Hynynen et al. 2002). Tree-level information is also important in the strategic planning of larger areas like municipalities, provinces or a country. Time series of large-scale DBH distributions are useful in the monitoring of changes in forested areas (Coomes and Allen 2007, Henttonen et al. 2019).

There are several probability density functions that have been used in the characterization of DBH distributions. The most common functions are Beta (Loetsch et al. 1973), Johnson SB (Hafley and Schreuder 1977) and Weibull (Bailey and Dell 1973). The two-parameter Weibull distribution has achieved popularity because of its convenient mathematical properties and flexibility to characterize various distribution shapes, such as right- and left-skewed, and Gaussian-shaped distributions (Bailey and Dell 1973). For the same reasons, we consider the Weibull distribution below.

The parameters of the Weibull distribution can be obtained using the parameter prediction method (PPM) (Kilkki et al. 1989) or the parameter recovery method (PRM) (Burk and Newberry 1984, Siipilehto and Mehtätalo 2011). Both methods originate from traditional field-based forest management inventories which today are vastly superseded by airborne laser scanning (ALS)-supported inventories following the area-based approach (ABA) (Næsset et al.

1997). ALS-supported inventories typically utilize field datasets which consist of sample plots with a full enumeration of DBHs (Næsset 2014, Maltamo and Packalen 2014).

PPM consists of three steps in ALS-supported forest inventories. First, Weibull distributions are fit to the DBH distributions of individual sample plots (e.g. Bailey and Dell 1973). Second, each of the estimated parameters is regressed against predictor variables such as ALS metrics (Gobakken and Næsset 2004) or forest attributes which were themselves predicted using ALS metrics (Packalén and Maltamo 2008). Finally, the model is applied to a wall-to-wall dataset of the predictor variables, typically consisting of grid cells with a similar size as the sample plots. An additional model for total stem frequency is needed to predict stem frequency per DBH class given the predicted Weibull parameters.

PRM consists of two steps. First, recovery attributes, such as basal area, mean DBH, stem frequency and moments/percentiles of the DBH distribution, are predicted for grid cells using ALS metrics. Then, the mathematical relationships between the recovery attributes and parameters of the Weibull distribution are used to create a non-linear system of equations. This system is solved using a root-finding approach such as Newton-Raphson (Siipilehto and Mehtätalo 2013). Because a numerical solution is not always possible when using PRM with predicted recovery attributes (Mehtätalo et al. 2007), we focus on PPM below.

Breidenbach et al. (2008) adapted a variety of PPM using a generalized linear model (GLM, see Cao 2004) for the prediction of DBH distribution in an ALS-supported forest inventory. Due to the structure of the field data (concentric sample plots), they used several truncated Weibull distributions conditional to specific DBH ranges. The key benefit of the approach compared to PPM is the estimation of the Weibull parameters in a single-step in which the observed DBHs are directly regressed against ALS metrics.

One challenge related to the parametric methods is that the basic formulations of the probability density functions cannot characterize multimodal DBH distributions (Zhang et al. 2001, Thomas et al. 2008). This is one of the reasons why the non-parametric nearest neighbor approach (NN) has been proposed for the prediction of DBH distributions in boreal ALS-based forest inventories, for example in Finland (Packalén and Maltamo 2008), Norway (Maltamo et al. 2009), and the United States (Mauro et al. 2019). NN enables the prediction of DBH distributions without any assumptions related to the shape of the distribution. A disadvantage of NN is that it cannot produce predictions that are beyond the range of training data. Therefore, the field data must always be comprehensive enough in order to avoid systematic errors in predictions (e.g. Breidenbach et al. 2012).

Only few studies have examined the estimation of DBH distributions at larger scales like municipalities, regions, or countries (Magnussen and Renaud 2016, Henttonen et al. 2019). The precision of direct (field-data based) estimates of forest attributes can be improved by means of remotely sensed data (e.g. McRoberts and Tomppo 2007, Haakana et al. 2020). In the context of DBH distributions, this has been investigated by Magnussen and Renaud (2016) who used a combination of forest inventory and ALS data to obtain model-assisted (MA) estimates of DBH distributions. They utilized multi-dimensional scaling (MDS) to link ALS data to the observed DBH distribution in four different study sites where the sample plot data were collected using stratified simple random sampling. The MDS approach is not scale dependent and it was applied both at the level of forest stand and at the level of stratum. From the point of view of boreal forests, the limitation of the approach proposed by Magnussen and Renaud (2016) might be the challenges associated with the prediction of stem frequency for small trees that grow under the dominant tree layer.

We used Norwegian national forest inventory (NFI) and ALS data for the modeling of DBH distributions. We also estimated the DBH distribution at the level of study area (henceforth “the study area DBH distribution") using the MA estimation and direct estimation. Our objectives were i) to predict DBH distributions at the plot-level using ALS metrics and ii) to study whether the MA estimates associated with the study area DBH distribution (by DBH classes) achieve higher precisions than the direct estimates. For objective i), we compared PPM, GLM, and NN. As a part of objective ii), we studied the efficiency of the estimators using both 2 cm and 6 cm DBH class widths.

# 2 MATERIAL

## 2.1 Study area

The 8.7 Mha study area (forest area 7.5 Mha) is approximately located between 58° N and 66° N in Norway. The study area was selected based on the coverage of the national ALS data (Figure 1). The study area contains significant climatic gradients caused by the large latitudinal extend. Significant elevational variations are also typical in the mountainous topography of Norway. The most economic value of forest is associated with the coniferous tree species: Norway spruce (*Picea abies* [L.] Karst.) and Scots pine (*Pinus sylvestris* [L.]). There are also several deciduous species growing in the area. Among the deciduous species, the most dominant are birch species (*Betula spp* [L.]) (Breidenbach et al. 2020b).

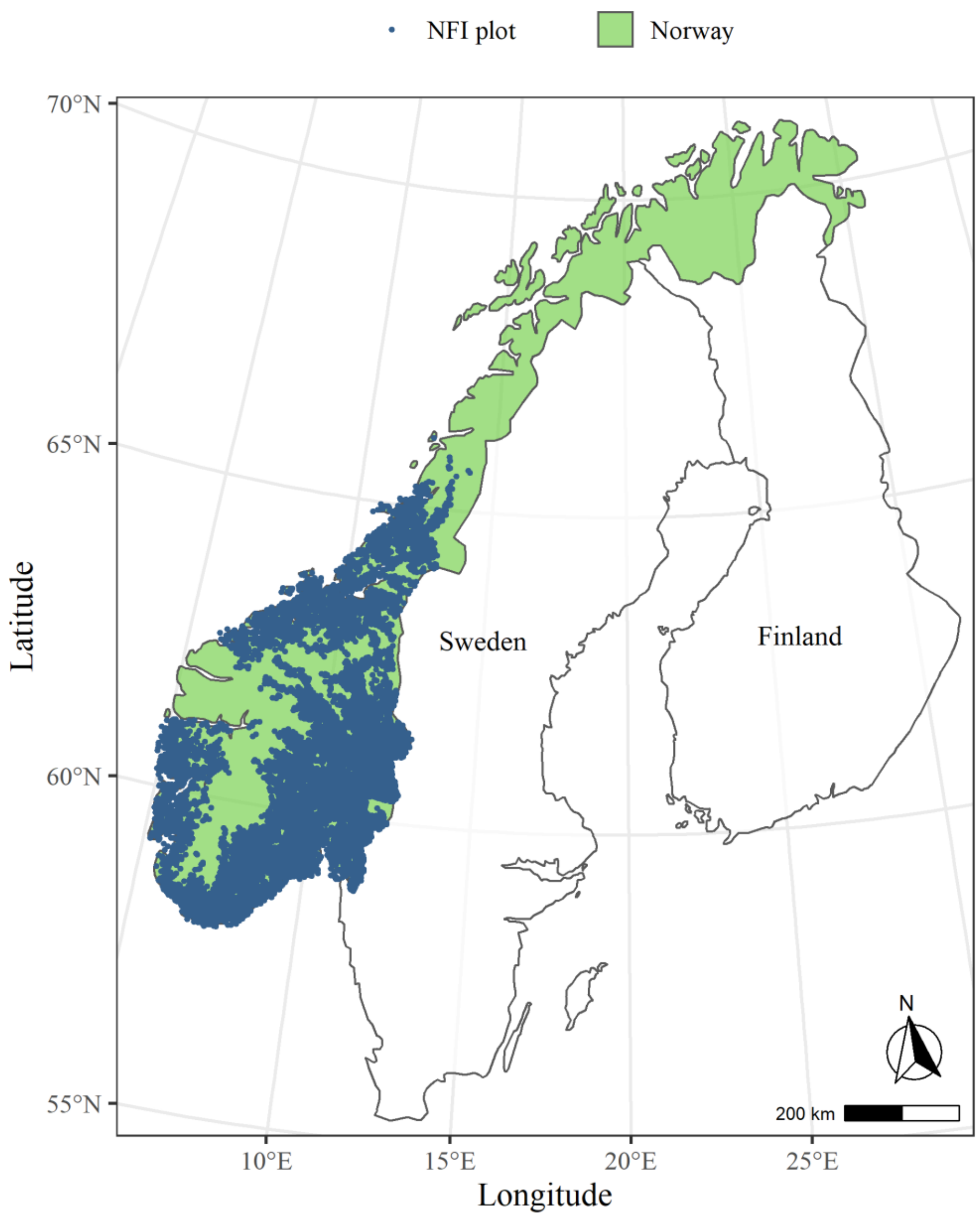

**Figure 1.** Approximate locations of the national forest inventory (NFI) field plots (n = 9 615) used in this study.

### 2.2 Field data

Our field dataset consisted of NFI plots measured between 2014 and 2018 from the lowland design stratum of the study area (Breidenbach et al. 2020a). The locations of the NFI plots follow a systematic grid with a resolution of 3 × 3 km, resulting in a total of 9 615 sample plots within the study area. Altogether, 8 384 of the sample plots were located within forest according to the NFI definition (10% crown cover and ability to reach 5 meters height). From the total number of forested plots, a total of 60 evident outliers resulting from harvests between field and ALS data acquisitions were excluded and assumed missing at random. The NFI plots are circular plots with an area of 250 $m^2$. In plots with forest cover, DBH and species were recorded for each tree with a DBH ≥ 5cm. For more detailed information related to the Norwegian NFI data, we refer to Breidenbach et al. (2020a).

Two datasets were created. The *modeling dataset* was used to compare PPM, GLM, and NN in the prediction of DBH distributions. The efficiency gains associated with the estimates, supported by using the best modeling approach, were analyzed using the *estimation dataset*.

The *modeling dataset* consisted of plots that were located in single-layered forests, because PPM and GLM utilize a unimodal probability density function. We categorized the NFI plots into three groups by dominant tree species (spruce, pine, and deciduous). The dominance of tree species was determined based on the species-specific timber volumes. Furthermore, we excluded plots with less than 5 measured trees, and plots that were located at the border of forest stands (split plots). The dataset consisted of 905, 813, and 259 plots in spruce, pine and deciduous dominated forests, respectively. Statistics associated with the forest attributes of the *modeling dataset* are presented in Table 1.

The *estimation dataset* consisted of all forested ($n_F$ = 8 324) and non-forested plots in the study area. The estimation dataset consisted of 3 156, 3 153 and 1 833 sample plots dominated by spruce, pine, and deciduous trees, respectively. The dominant tree species was undefined in a total of 182 forested plots without measured trees (young forests with no measured trees). The characteristics of the *estimation dataset* (plots within forest) are shown in Table 1 and Figure 2.

**Table 1.** Characteristics of selected forest attributes in the modeling dataset and the estimation dataset. V – volume, G – basal area, N – stem frequency, DBH – diameter at breast height, DG – basal area weighted mean DBH, Hg – Lorey's height.

| Statistic | Dominant tree species | Modeling dataset | | | | Estimation dataset | | | |
|---|---|---|---|---|---|---|---|---|---|
| | | Mean | Sd | Min | Max | Mean | Sd | Min | Max |
| V ($m^3 \cdot ha^{-1}$) | Spruce | 245.3 | 153.5 | 6.6 | 1000.4 | 146.2 | 141.1 | 0 | 1000.4 |
| | Pine | 163.6 | 93.0 | 11.4 | 615.1 | 103.1 | 95.6 | 0 | 702.2 |
| | Deciduous | 100.1 | 72.9 | 4.2 | 473.4 | 68.6 | 79.2 | 0 | 680.3 |
| G ($m^2 \cdot ha^{-1}$) | Spruce | 30.1 | 13.3 | 1.6 | 96.2 | 19.6 | 14.4 | 0 | 96.2 |
| | Pine | 22.5 | 9.8 | 2.6 | 59.6 | 15.2 | 11.3 | 0 | 79.1 |
| | Deciduous | 17.7 | 9.7 | 1.1 | 60.6 | 12.0 | 11.0 | 0 | 69.7 |
| N ($ha^{-1}$) | Spruce | 1262 | 647 | 200 | 4560 | 1030 | 698 | 0 | 5000 |
| | Pine | 785 | 481 | 200 | 4200 | 662 | 522 | 0 | 4520 |
| | Deciduous | 1336 | 863 | 200 | 5560 | 1012 | 839 | 0 | 5560 |
| DG (cm) | Spruce | 23.0 | 6.2 | 8.5 | 47.0 | 20.3 | 7.5 | 5.0 | 60.6 |
| | Pine | 25.2 | 6.0 | 10.7 | 49.3 | 23.7 | 7.6 | 5.2 | 70.0 |
| | Deciduous | 17.5 | 6.4 | 7.1 | 51.3 | 15.6 | 7.2 | 5.0 | 155.0 |
| Hg (m) | Spruce | 16.0 | 4.0 | 5.6 | 28.8 | 13.4 | 4.8 | 3.2 | 32.3 |
| | Pine | 14.5 | 3.3 | 6.6 | 27.0 | 12.3 | 4.2 | 2.3 | 70.1 |
| | Deciduous | 10.9 | 3.5 | 4.9 | 26.0 | 9.8 | 4.1 | 2.7 | 26.6 |
| Tree-level DBH (cm) | Spruce | 15.4 | 8.2 | 5.0 | 79.8 | 13.4 | 7.9 | 5.0 | 79.8 |
| | Pine | 16.9 | 9.0 | 5.0 | 68.0 | 14.6 | 8.9 | 5.0 | 96.0 |
| | Deciduous | 11.5 | 6.0 | 5.0 | 58.0 | 10.7 | 6.0 | 5.0 | 155.0 |

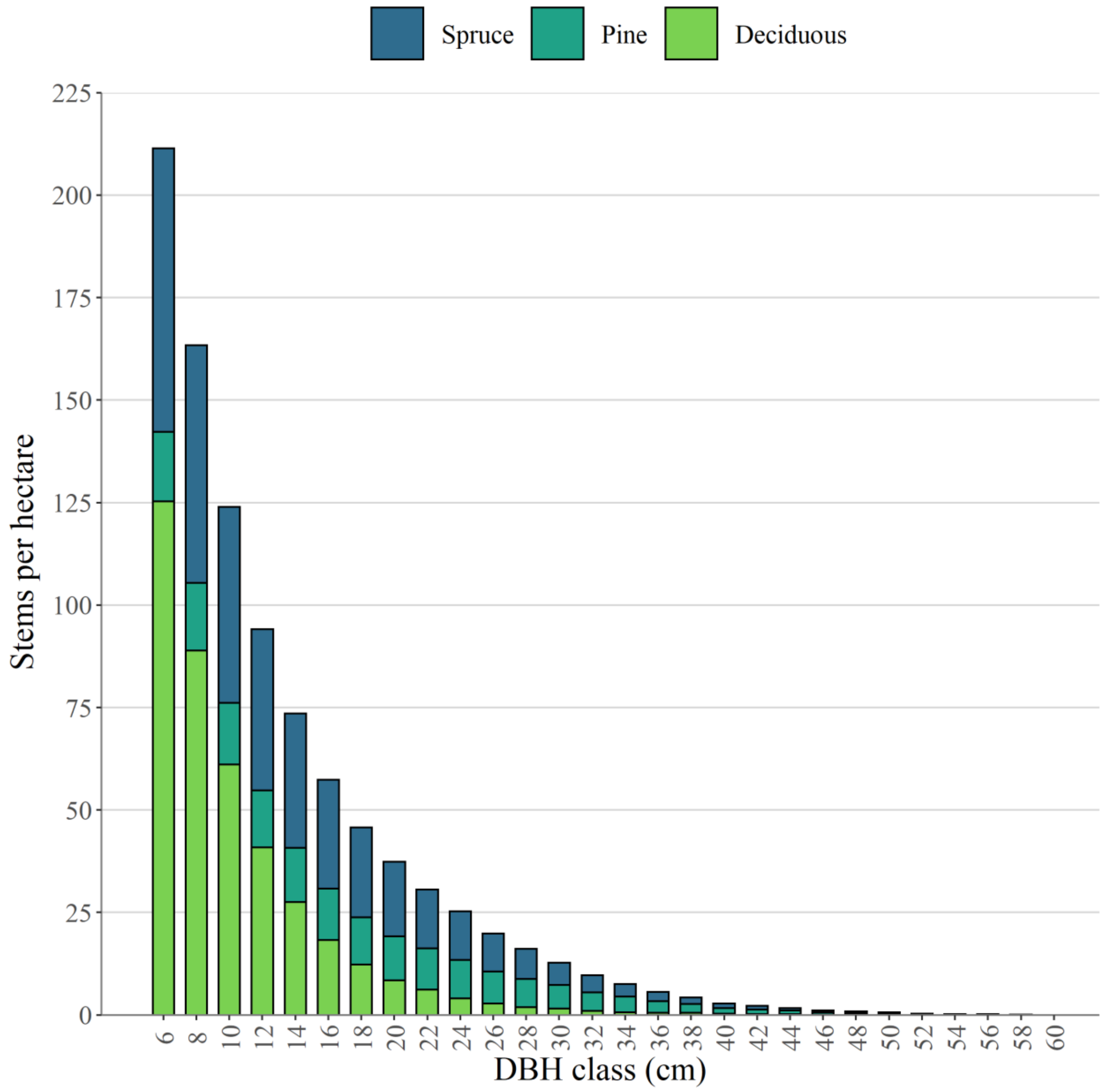

**Figure 2.** DBH distribution associated with the estimation dataset.

### 2.3 Airborne laser scanning data and extracted metrics

The acquisition of ALS data was carried out between 2010 and 2018. The study area was covered by several ALS campaigns, which means that the data acquisition parameters differ across the study area. The mean pulse density varied between 2–5 points per square meter among the ALS campaigns. A digital terrain model (DTM, 1 × 1 m) was created using the last returns of the ALS datasets (Kartverket 2019). The height measurements of the ALS datasets were normalized to above ground heights by subtracting the DTM elevation from the orthometric ALS height measurements.

We extracted height, intensity, and echo proportion metrics from the ALS data for the sample plots. The height metrics consist of minimum, maximum, mean, variance and coefficient of variation, skewness and kurtosis, as well as height percentiles and densities. For density metrics, the height range from the ground to the 95% percentile was divided into 10 height slices of equal size (starting from slice 0 which is the closest to the ground level). The densities were computed as the proportion of returns above the height slices 0, 2, 4, 6, 8, and 9 to all returns (without echo categorization). In addition, the proportion of echoes above 2 meters was computed. The intensity metrics were computed based on the intensity recordings of ALS data and consisted of variance, coefficient of variation, and ratio between the mean of ground echo intensities and the mean of vegetation echo intensities. The metrics were computed by echo categories: first, last, and all. We extracted a total of 23 metrics from the ALS data per echo category. Table 2 shows the extracted metrics and their abbreviations.

**Table 2.** Metrics extracted from the ALS data for the sample plots. The metrics were computed by echo categories first, last, and all.

| Abbreviation | Description |
|---|---|
| *Height metrics* | |
| hmin, hmax, hmean, hvar, hcv | Minimum, maximum, mean, variance, coefficient of variation |
| hskew, hkurt | Skewness and kurtosis |
| h10, h25, h50, h75, h90, h95 | Height percentiles for 10%, 25%, …, 95% |
| d0, d2, d4, d6, d8, d9 | Height densities |
| *Intensity metrics* | |
| ivar, icv, igratio | Variance, coefficient of variation, ratio between the mean of ground echo intensities and the mean of vegetation echo intensities |
| *Proportion metrics* | |
| proph | Proportion of echoes above 2 meters |

# 3 METHODS

## 3.1 DBH distribution models

### *3.1.1 Parameter prediction method using linear mixed-effects models (PPM)*

We applied a left-truncated two-parameter Weibull function (Zutter et al. 1986) in the modeling of DBH distributions. The left-truncation was used since the trees with a DBH < 5 cm were not measured. The truncated two-parameter Weibull probability distribution is as follows:

$$f(x|\ a,\ b)\ =\ \begin{cases} \frac{a}{b}\left(\frac{x}{b}\right)^{a-1} e^{\left[\left(\frac{T}{b}\right)^{a}-\left(\frac{x}{b}\right)^{a}\right]}, x \geq T \\ 0\ , x < T \end{cases} \qquad (1)$$

where x is DBH, T is the fixed left-truncation point (5 cm), a is the shape parameter, and b (T < b) is the scale parameter.

The estimates of Weibull parameters are needed for each sample plot in order to fit regression models for the scale and shape parameters. We estimated the parameters of Weibull distribution by maximizing the log-likelihood function. The log-likelihood function for plot j is maximized given the Weibull parameters (a and b)

$$\log(L_j) = \sum_{k=1}^{n_j} \log(f(x_k|a, b)) \tag{2}$$

where $n_j$ is the stem frequency in plot j and $f(.)$ is the truncated Weibull probability distribution (Eq. 1) and $x_k$ refers to the observed DBH in plot $j$. This approach is referred to as the maximum likelihood (ML) method (Bailey and Dell 1973, Mehtätalo and Lappi 2020 p. 337–338). We used the *mle* function of the package *stats4* in *R* (R Core Team 2020) for the ML estimation of the plot-level Weibull parameters.

Because the field data have a hierarchical structure (NFI plots within ALS project), we fitted linear mixed-effects models with random intercepts for the shape and scale parameter of the Weibull distribution. The coefficients were estimated using the restricted maximum likelihood approach (e.g. Fahrmeir et al. 2013, p. 109), and the models were restricted to the p=3 most important predictor variables

$$y_{lij} = b_{li} + \beta_{l0} + \beta_{l1}x_{lij}^{(1)} + \cdots + \beta_{lp}x_{lij}^{(p)} + \epsilon_{lij} \tag{3}$$

where $y_{lij}$ is the response variable l={shape (a), scale (b)} predicted at plot j using Eq. (2) in ALS project i, $\beta_{l0} + \beta_{l1}x_{lij}^{(1)} + \cdots + \beta_{lp}x_{lij}^{(p)}$ is the fixed part of the model in which the $\beta s$ are the coefficients to be estimated and $x_{lij}^{(p)}$ are predictor variables, $b_{li} + \epsilon_{lij}$ is the random part of the model in which $b_{li} \sim N(0,\ \sigma_b^2)$ represents the random part of the intercept in ALS project i and $\epsilon_{lij} \sim N(0, \sigma^2)$ represents the residual error of plot j in project i. We used the *nlme* package (Pinheiro et al. 2020) for the estimation of the model parameters. Ultimately, we transformed the predicted Weibull distributions to DBH distributions (i.e. histograms) using a bin width of 2 cm. The transformation was based on the cumulative Weibull distribution function and total stem frequencies observed in the sample plots. Because of the small number of trees with a DBH > 50 cm (Figure 2), we consider DBH classes with the mid-points in the range of 6–50 cm.

*3.1.2 Parameter prediction using generalized linear models (GLM)*

A GLM-like framework (Cao 2004; Mehtätalo and Lappi 2020, p. 370–373), simply denoted as GLM in the following, was used to construct parameter models that directly link the measured DBHs and ALS metrics (Breidenbach et al. 2008). We applied the truncated Weibull distribution described in Eq. 1 and maximized the likelihood function

$$\log(L) = \sum_{j=1}^{n} \log(L_j) \tag{4}$$

for the Weibull parameters $\theta_{lijk}$={b, c} by optimizing the parameters of the regression function

$$\theta_{lijk} = b_{li} + b_{lij} + \beta_{l0} + \beta_{l1} x_{lijk}^{(1)} + \cdots + \beta_{lp} x_{lijk}^{(p)} \tag{5}$$

where n is the number of plots, $b_{li}$ is a random effect at the ALS-project level and $b_{lij}$ is a plot-level random effect to consider the dependence among trees within a sample plot since they share the same predictor variables. The log link function was used in the modeling of the Weibull parameters.

The models for shape and scale parameters were simultaneously fitted by maximizing the penalized likelihood. We fitted the models using the *gamlss* R package (Rigby and Stasinopoulos 2005). We dropped the project-level random effect ($b_{li}$) describing the grouping of plots within the ALS projects after preliminary analysis because of numeric instability.

Ultimately, we transformed the predicted Weibull distributions to DBH distributions (i.e. histograms) as described for PPM.

### *3.1.3 Nearest neighbor approach*

While NN is considered non-parametric, the number of nearest neighbors (k), the number of predictor variables (p), the response configuration, and a distance metric need to be chosen. After preliminary analysis, we set k=p=5 and used basal area, stem frequency, basal area, basal area weighted mean DBH, and Lorey's height as the response configuration. We used the most similar neighbor (MSN) distance, which is based on the canonical correlation analysis between the response and predictor variables (Moeur and Stage 1995). The squared MSN distance between reference u and target j observation derived from canonical correlation analysis is as follows:

$$d_{uj}^2 = \underset{1 \times p}{(x_u - x_j)} \underset{p \times p}{\Gamma \Lambda^2 \Gamma'} \underset{p \times 1}{(x_u - x_j)'} \quad (6)$$

Where $d_{uj}^2$ is the squared MSN distance, $x_u$ and $x_j$ are row vectors of predictor variables for training and target plots, $\Gamma$ is a matrix of canonical coefficients of predictor variables, and $\Lambda^2$is a diagonal matrix of squared canonical correlations.

The inverse of the distances $d_{uj}^2$ was used to weight the reference tree lists for the target observations and the tree lists were transformed to DBH distributions (i.e. histograms). The NN-based DBH distributions represent the absolute stem frequency which is an advantage compared to PPM and GLM which model relative stem frequencies. However, in order to compare the performance of NN with PPM and GLM in the prediction of DBH distributions, we converted the NN-based DBH distributions to be consistent with the observed stem frequencies. In the MA estimation, the absolute NN-based stem frequencies were used. NN was carried out using the *yaImpute* R package (Crookston and Finley 2007). The NN-based DBH distribution usually results in spikes and pits (Strunk et al. 2017). Therefore, we smoothed the NN-based DBH distributions with a 3-bin moving average.

*3.1.4 Selection of predictor variables*

We used the same predictor variables in PPM and GLM, which enabled us to trace differences in the predictive performance. The predictor variable selection was carried out based on PPM and comprised two steps: an automatized step using an optimization algorithm and a manual step for the final evaluation of model fit.

The automatized step was based on the heuristic optimization algorithm, known as simulated annealing (Kirkpatrick et al. 1983, Packalén et al. 2012), which minimizes a cost function by repeatedly fitting the model and randomly changing the combination of predictor variables. The initial temperature, the number of iterations per temperature, and a cooling factor determined the number of iterations. The cooling factor (value < 1) is used to cool the system by multiplying the current temperature value. The temperature value determines the probability to accept a worse solution. The cost function was the root-mean-squared error associated with the response variable.

The automatized step was used to reduce the number of predictor variable candidates. The manual step included visual and numerical assessments related to the model fit statistics. Finally, three predictor variables were selected for each Weibull parameter.

In case of NN, we only applied the automatized step. We repeated the selection of predictor variables five times in order to observe the fluctuation in the ultimate cost value caused by the heuristic optimization. We selected the combination of predictor variables that resulted in the smallest cost value. The number of predictor variables was five.

*3.1.5 Evaluation of the predicted DBH distributions*

We predicted the DBH distributions using leave-one-out cross validation. The predictive performances of the DBH distribution models were evaluated by the DBH classes. The errors associated with DBH class $c$ were evaluated using root-mean-square error ($RMSE_c$, Eq. 7) and mean difference ($MD_c$, Eq. 8).

$$RMSE_c = \sqrt{\frac{\sum_{j=1}^{n}\left(y_{cj} - \hat{y}_{cj}\right)^2}{n}} \tag{7}$$

$$MD_c = \frac{\sum_{j=1}^{n}\left(y_{cj} - \hat{y}_{cj}\right)}{n} \tag{8}$$

where $y_j$ and $\hat{y}_j$ are the observed and predicted stem frequency associated with DBH class c = {6, 8 ,…, 50} in plot j, and $n$ is the total number of sample plots in the dominant tree species group (pine, spruce, or deciduous).

We used a variant of the error index (Reynolds et al. 1988) to assess the predicted DBH distribution. The error index (EI, Eq. 9) measures the goodness of the DBH distribution model at the level of all sample plots. This means that systematic prediction errors will be retained in the EI values whereas randomly occurring over- and under predictions may cancel out.

$$EI = \sum_{c=1}^{k}\left|\frac{\sum_{j=1}^{n} y_{cj}}{n} - \frac{\sum_{j=1}^{n} \hat{y}_{cj}}{n}\right| \tag{9}$$

where k is the number of DBH classes.

## 3.2 Estimation of the study area DBH distribution

### *3.2.1 Direct estimation of the DBH distribution*

Direct estimators only utilized the NFI sample for the estimation. The direct estimators were used as references in the comparison with MA estimators. The direct estimator for the mean stem frequency per forest land hectare ($\hat{\mu}$) within a given DBH class is

$$\hat{\mu} = \frac{\sum_{j \in S} y_j I_j}{\sum_{j \in S} I_j} \tag{10}$$

where $y_j$ is the observed stem frequency per hectare forest land at sample plot j belonging to sample S. The binary indicator variable $I_j$ is 1 for forested plots and 0 for non-forest plots, respectively.

A variance estimator assuming simple random sampling (SRS) is frequently used in NFIs, because there is no design-unbiased variance estimator available for systematic sampling. In systematic sampling, the SRS estimator is under practical conditions basically always conservative and thus overestimates the variance (e.g., Magnussen et al. 2020).

It is, however, possible to reduce the overestimation of the variance by applying local difference estimators instead of the SRS estimator (Magnussen et al. 2020, Räty et al. 2020, Heikkinen 2006). The local difference estimator proposed by Grafström and Schelin (2014) was therefore used in this study (Räty et al. 2020). The Grafström-Schelin (GS) variance estimator utilizes the local differences of sample plots in a neighborhood $S_j^*$ which is a subset of sample $S$ and comprises plot j and its closest neighbors in the four cardinal directions.

$$\widehat{Var}_{GS}(\hat{\mu}) = \frac{1}{(\sum_{j \in S} I_j)^2} \sum_{j \in S} \frac{n_j^*}{n_j^* - 1} \left( z_j - \frac{1}{n_j^*} \sum_{k \in S_j^*} z_k \right)^2 \quad (11)$$

where $z_j = y_j - \hat{\mu} I_j$, and $n_j^*$ is the number of sample plots in neighborhood $S_j^*$.

The standard error (SE) of the estimate is

$$SE(\hat{\mu}) = \sqrt{\widehat{Var}(\hat{\mu})} \quad (12)$$

Because of its status as a quasi-standard, we present the variance estimator assuming SRS in Appendix 1.

*3.2.2 Model-assisted estimation of the DBH distribution*

The MA estimator is used to include the predicted forest attributes in the estimation process. In case a good model exists, this can reduce the variance in comparison to the direct estimator and thus increase the efficiency of the estimator (Särndal et al. 1992). The MA estimator for mean stem frequency per forest land hectare associated with a DBH class is

$$\hat{\mu}_{MA} = \tilde{\mu}_S + \hat{\mu}_{cor} \quad (13)$$

where $\tilde{\mu}_S$ is the synthetic estimate of the mean stem frequency per unit forest land based on the aggregation of a wall-to-wall map and $\hat{\mu}_{cor} = \frac{\sum_{j \in S} e_j}{\sum_{j \in S} I_j}$ is the correction factor with $e_j = y_j - \hat{y}_j$ as the model residual associated with plot j. The correction factor ($\hat{\mu}_{cor}$) adjusts possible systematic errors caused by the models. The variance of $\hat{\mu}_{MA}$ was estimated using the GS estimator (Eq. 11) with

$$z_j = e_j - \bar{e} \quad (14)$$

in which $\bar{e}$ is the mean of residuals.

*3.2.3 Application of the MA estimator*

We used leave-one-out cross validated predictions $\hat{y}$ in the MA estimation. The MA estimator was applied using the forest/non-forest information and tree species observed in the sample plots to study the theoretical benefit of using the DBH distribution models. However, this information is only known in the sample plots. A predicted tree species map that also contains a non-forest class (i.e. forest/non-forest information), is used in the operational mapping of stem frequency. The tree species map introduces further uncertainty in the estimation process. Therefore, we utilized the predicted tree species map (prediction for the sample plots) in order to study the MA estimation from the viewpoint of operational practices.

The forest/non-forest information had an overall accuracy (OA) of 92% while the tree species predictions had an OA 74% and 90% at the plot-level and stand-level, respectively (Breidenbach et al. 2020b). An error associated with the forest/non-forest class of the map has the effect that a forested plot is treated as a non-forest plot ($|e_j|$ equals to the observed stem frequency) or vice versa ($|e_j|$ equals to the predicted stem frequency). An error in the tree species class of the map has the effect that a suboptimal model (NN models by dominant tree species) is used for the prediction of the DBH distribution.

*3.2.4 Evaluation of the estimators*

In order to compare the efficiency of the direct and MA estimators, we used the half width of 95 % confidence intervals, correction factors ($\hat{\mu}_{cor}$), and the relative efficiency (RE)

$$RE = \frac{\widehat{Var}(\hat{\mu})}{\widehat{Var}(\hat{\mu}_{MA})}. \tag{15}$$

RE values larger than 1.0 indicate that the MA estimator of variance results in a smaller variance value than the direct estimator of variance. Assuming SRS, $RE \times n$ is the number of sample plots required to obtain the same variance as the MA estimate. For example, RE = 1.5 means that the direct estimator results in the same variance as the MA estimator given that the number of sample plots is increased by 50%.

# 4 RESULTS

## 4.1 Modeling of DBH distributions

In this section, we consider the goodness of the predicted DBH distributions from the viewpoint of large-scale estimation. The predictor variables selected for PPM, GLM and NN, and the estimated model parameters associated with the PPM and GLM models are presented in Appendix 2 (Tables A2.1, A2.2 and A2.3).

NN outperformed PPM and GLM in terms of the EI values associated with the predicted DBH distributions in the pine- and spruce-dominated plots (Table 3). GLM outperformed PPM regardless of the dominant tree species and, for the deciduous-dominated plots, the GLM also resulted in a slightly smaller EI value than NN. The mean of EI values over all tree species groups (weighted by the number of plots) were 184, 150, and 47 stems per hectare for PPM, GLM, and NN, respectively. GLM and PPM resulted in systematic errors in certain DBH classes whereas NN produced moderate MD values for all DBH classes (Figure 3). The smaller systematic error is the reason for the smaller EI values associated with NN compared to the other approaches.

**Table 3.** Error index (EI) values associated with the predicted DBH distribution at the level of plots dominated by spruce, pine and deciduous tree species.

| | Spruce-dominated plots, EI (stems per hectare) | Pine-dominated plots, EI (stems per hectare) | Deciduous-dominated plots, EI (stems per hectare) |
|---|---|---|---|
| PPM | 170.9 | 197.5 | 193.8 |
| GLM | 153.6 | 162.8 | 100.8 |
| NN | 40.6 | 34.0 | 108.9 |

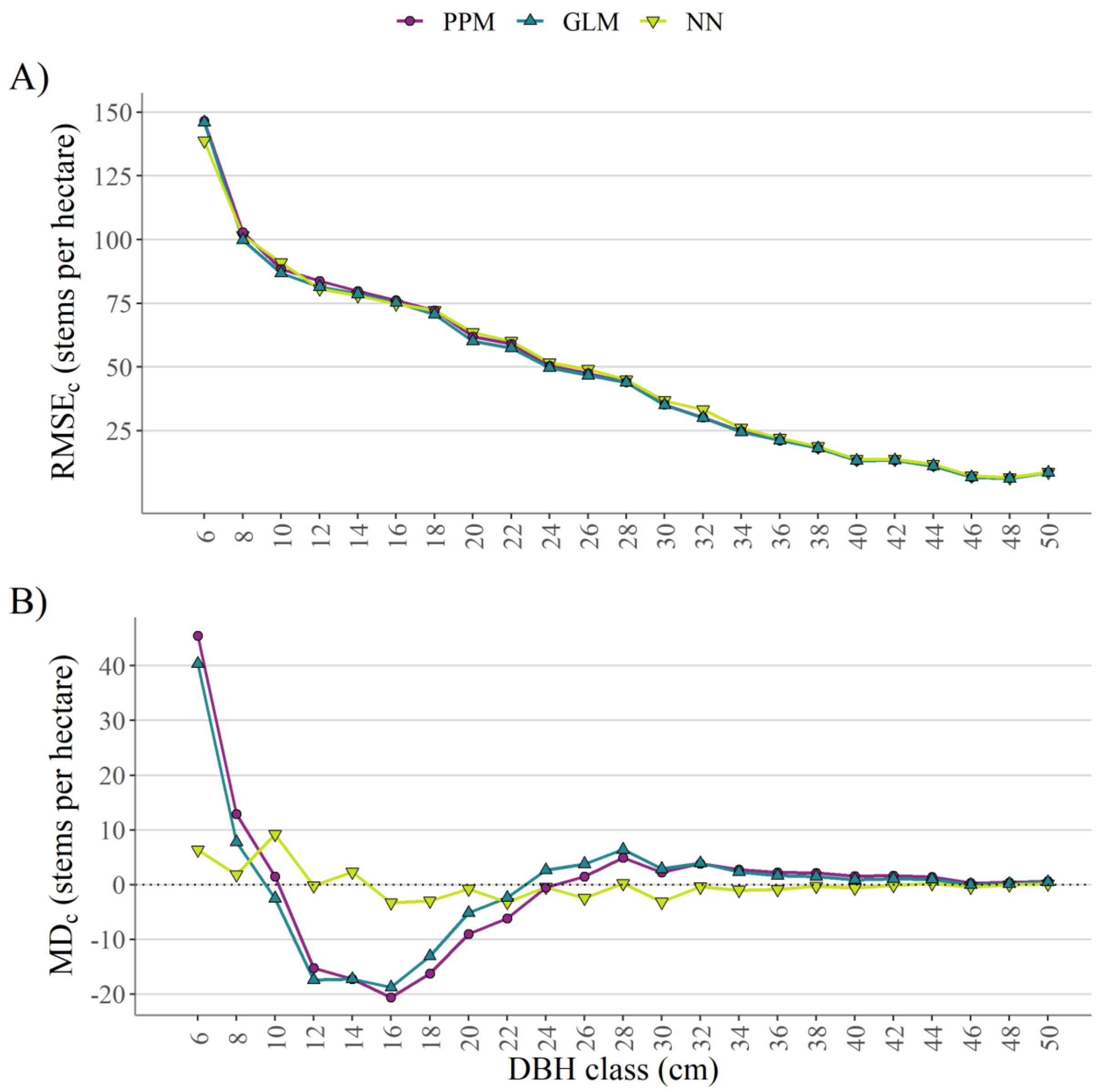

**Figure 3**. A) Root mean squared error and B) mean difference associated with the predicted stem frequencies of the DBH classes (the subscript c refers to DBH class) in the spruce-dominated plots (n = 905). PPM – parameter prediction method using linear mixed-effects models, GLM – parameter prediction using generalized linear models, NN – Nearest neighbor approach

## 4.2 Estimation of the study area DBH distribution

We used NN in the MA estimation of DBH distribution for the study area due to its better predictive performance, compared to PPM and GLM, especially with respect to systematic errors. The comparison of the DBH distribution models (section 4.1) was carried out with the *modeling dataset* that did not include multi-layered forests, plots with less than 5 trees, or split plots. Because the models presented in section 4.1 are suboptimal for the MA estimation of the study area DBH distributions, we trained new NN models for each dominant tree species using the *estimation dataset*. These NN models were used in the MA estimation of the study area DBH distribution. The evaluation of the NN models by DBH classes is presented in Figure 4. Figure 4A shows that the RMSE values were larger than in the comparison of modeling approaches (Figure 3A), which is to be expected when including plots representing more heterogenous conditions. The MD values (Figure 4B) were, however, still moderate as observed in section 4.1 (Figure 3B).

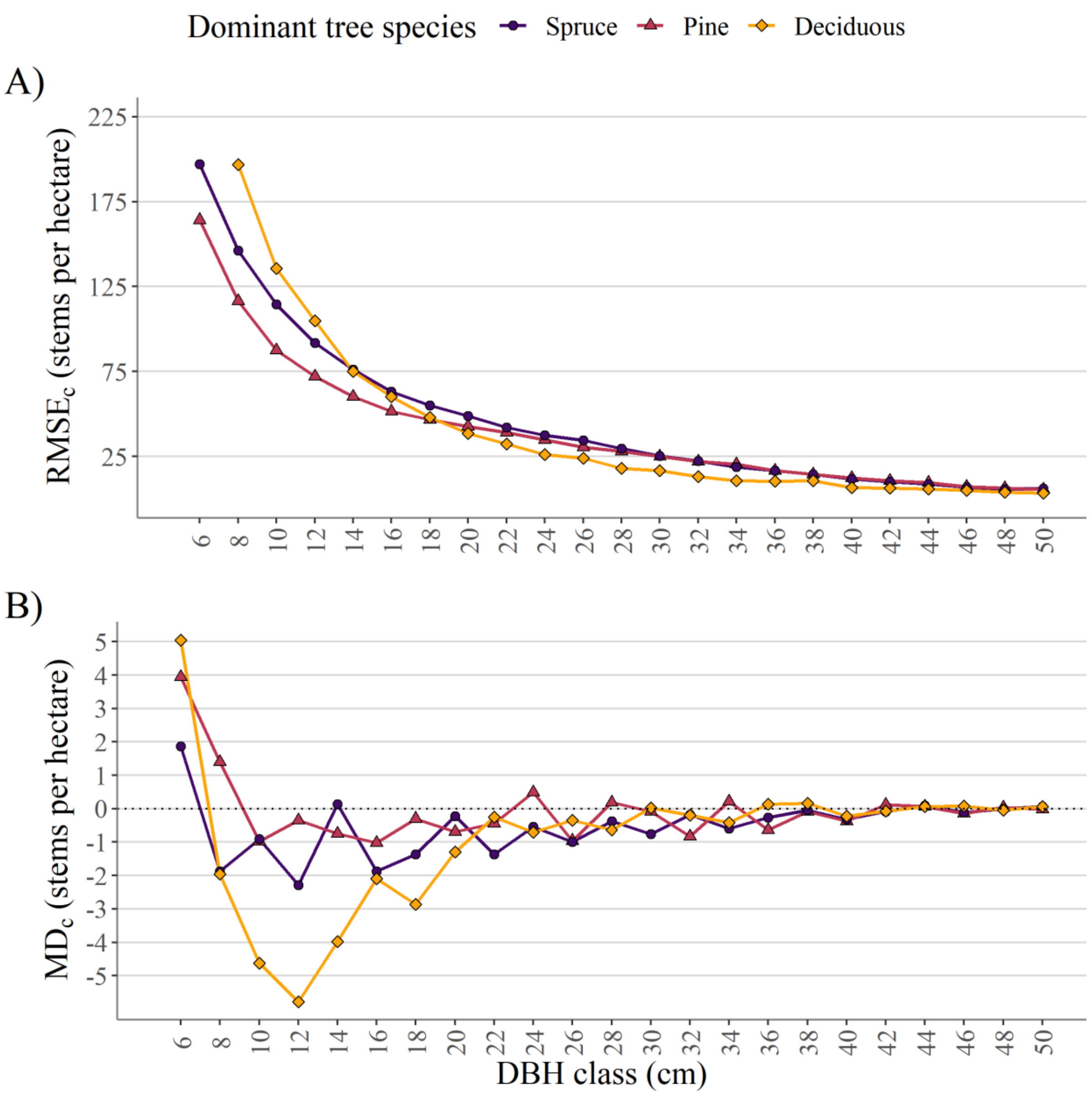

**Figure 4.** Evaluation of the nearest neighbor models by dominant species used in the model-assisted estimation of the study area DBH distribution. A) root mean squared error and B) mean difference by DBH classes (the subscript c refers to DBH class).

The MA estimator was more efficient than the direct estimator in the majority of the DBH classes regardless of the width of DBH class (Figure 5). The estimates on the left tail of the DBH distribution profited more from the use of remotely sensed data than those at the right tail of the DBH distribution. Furthermore, the RE values were larger with 6 cm than 2 cm DBH classes and smaller than 1.00 in the DBH classes with a mid-point ≥ 40 cm. Uncertainties that propagated from the tree species map reduced the RE values associated with the DBH classes up to 10 %, but the RE values were nonetheless ≥ 0.95 for all DBH classes (Figure 5). The RE values associated with the MA estimation of the total stem frequency were 2.11, given the observed forest/non-forest and tree species information, and 1.66, given the predicted tree species map.

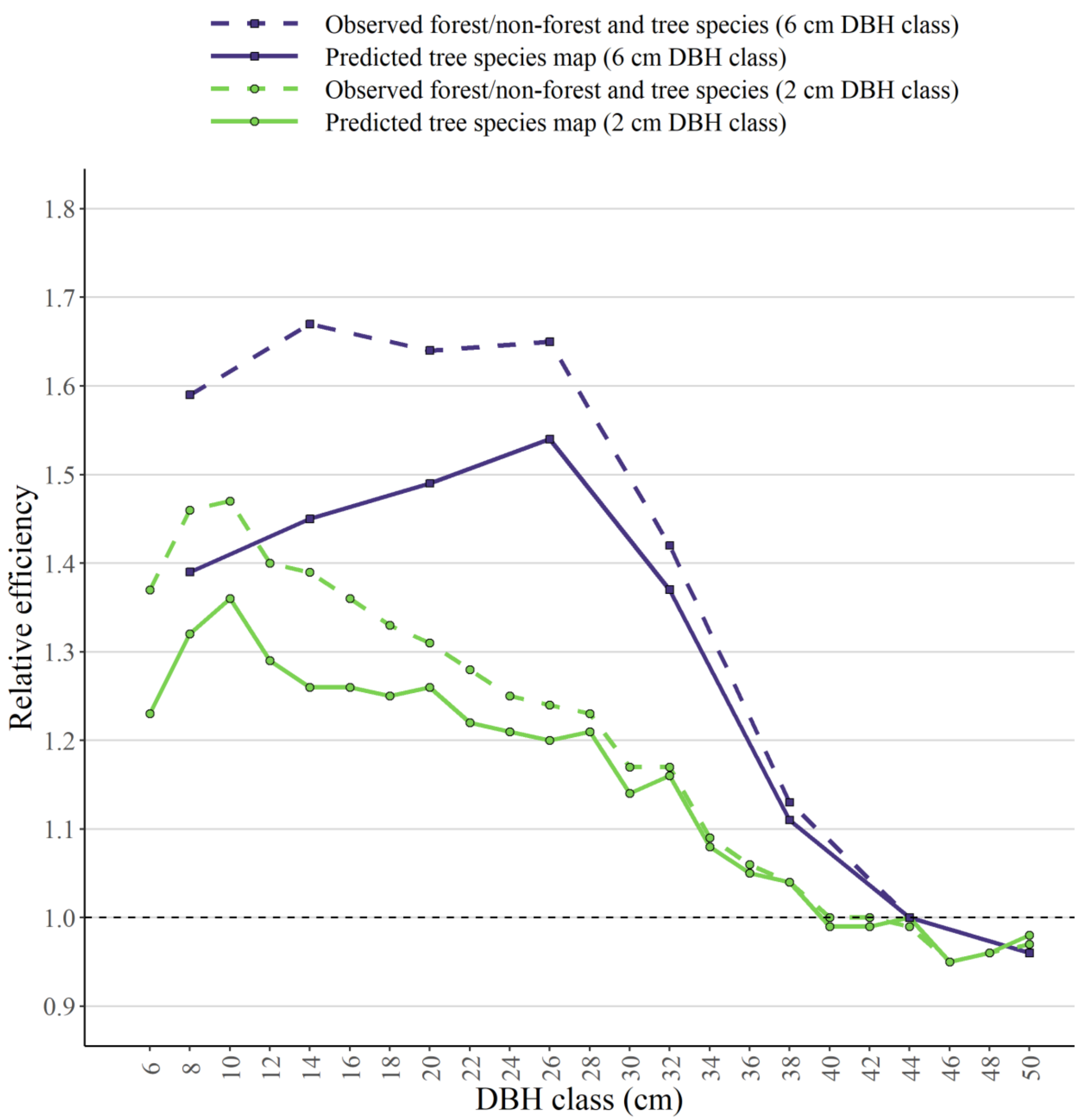

**Figure 5.** Relative efficiency (RE) of the model-assisted estimator by DBH classes.

The confidence intervals and correction factors associated with the study area DBH distribution are presented in Table 4. The non-zero correction factors mean that systematic errors existed both with the observed forest/non-forest information and tree species, and with the tree species map. As can be expected, the correction factors were generally larger when the tree species map was used instead of the observed information. The correction factor was also required (non-zero) for the estimate of total stem frequency.

Results associated with the SRS variance estimator, are given in Appendix 1.

**Table 4.** Characteristics associated with the direct and model-assisted estimates of the DBH distribution at the level of the study area. Variances were estimated using the Graftström-Schelin (GS) estimator. The model-assisted estimates were computed using the observed tree species and predicted tree species map (in parenthesis). Values in per-cent are given in relation to the direct estimate.

| DBH class (cm) * | Direct estimate $\hat{\mu}$ (stems per hectare) | Half 95% $CI_{GS}$ $\hat{\mu}$, % | Half 95% $CI_{GS}$ $\hat{\mu}_{MA}$, % | Correction factor $\hat{\mu}_{cor}$, % | DBH class (cm) * | Direct estimate $\hat{\mu}$ (stems per hectare) | Half 95% $CI_{GS}$ $\hat{\mu}$, % | Half 95% $CI_{GS}$ $\hat{\mu}_{MA}$, % | Correction factor $\hat{\mu}_{cor}$, % |
|---|---|---|---|---|---|---|---|---|---|
| 6 | 206.74 | 2.57 | 2.20 (2.32) | 3.16 (9.54) | 30 | 12.41 | 4.46 | 4.11 (4.17) | -2.46 (0.41) |
| 8 | 159.83 | 2.41 | 2.00 (2.10) | 0.81 (6.63) | 32 | 9.49 | 5.10 | 4.72 (4.73) | -4.54 (-1.17) |
| 10 | 121.24 | 2.39 | 1.97 (2.05) | -0.57 (4.32) | 34 | 7.38 | 5.57 | 5.33 (5.37) | -3.23 (0.71) |
| 12 | 92.04 | 2.46 | 2.08 (2.16) | -1.88 (2.22) | 36 | 5.52 | 6.28 | 6.12 (6.13) | -5.79 (-0.83) |
| 14 | 71.97 | 2.46 | 2.09 (2.20) | -1.03 (2.13) | 38 | 4.20 | 7.22 | 7.10 (7.09) | -0.11 (3.95) |
| 16 | 56.08 | 2.62 | 2.25 (2.34) | -2.41 (0.29) | 40 | 2.75 | 8.71 | 8.71 (8.74) | -11.93 (-7.79) |
| 18 | 44.70 | 2.79 | 2.42 (2.50) | -2.54 (-0.55) | 42 | 2.17 | 9.83 | 9.85 (9.86) | -0.16 (1.86) |
| 20 | 36.59 | 3.03 | 2.65 (2.71) | -1.42 (0.09) | 44 | 1.66 | 11.30 | 11.38 (11.29) | 4.73 (7.11) |
| 22 | 29.95 | 3.21 | 2.84 (2.91) | -2.20 (-0.50) | 46 | 1.04 | 13.37 | 13.68 (13.69) | -7.67 (-4.71) |
| 24 | 24.70 | 3.40 | 3.04 (3.10) | -0.42 (1.35) | 48 | 0.80 | 15.02 | 15.34 (15.33) | 0.65 (5.11) |
| 26 | 19.38 | 3.86 | 3.46 (3.52) | -4.16 (-1.59) | 50 | 0.60 | 18.98 | 19.28 (19.20) | 5.16 (8.12) |
| 28 | 15.78 | 4.12 | 3.71 (3.75) | -1.18 (1.16) | All | 928.23 | 1.57 | 1.08 (1.21) | 1.40 (4.23) |

* DBH class refers to the middle of DBH class.

## 5 DISCUSSION

We compared PPM, GLM and NN for the modeling of DBH distributions at the plot-level and selected the best approach for the MA estimation of the study area DBH distribution. While GLM outperformed PPM in the prediction of DBH distributions, GLMs have not achieved popularity in the prediction of DBH distributions in the ALS-based forest inventories so far. Breidenbach et al. (2008) predicted DBH distributions using GLMs and ALS data but they did not consider the clustered structure of the data. Because trees observed in one plot are not independent of each other, the standard errors obtained by Breidenbach et al. (2008) were therefore too small. This issue was resolved here by adding a plot-level random effect.

The advantage of GLM is that it is a single-step approach and supersedes the estimation of Weibull parameters per plot required in PPM. In order to reasonably estimate the Weibull parameters at the plot level, numerous DBH recordings are required. In some cases, the lack of trees may be an issue with relatively small plot sizes (e.g. 250 $m^2$) which are often used in the ALS-based forest inventories. The predictive performance of GLM may also be improved by additional smoothing terms resulting in a generalized additive model (Rigby and Stasinopoulos 2005). In our preliminary analysis, smoothing terms did not, however, improve the predictive performance of the models.

An alternative to the parametric approaches is NN which predicts DBH distributions without any assumptions regarding the shape of DBH distribution. We showed that NN clearly outperformed the parametric approaches, especially among the coniferous species which are of the most economic interest in Nordic timber production. Our findings are in line with Packalén and Maltamo (2008) who reported that NN outperformed PPM in the prediction of DBH distributions for managed boreal forest stands. They utilized predicted forest attributes as predictor variables in the parameter models and focused on the prediction of species-specific DBH distributions.

In the selection of the modeling approach for DBH distributions, it is worth noting that the shapes of DBH distributions are affected by silvicultural activity (Rouvinen and Kuuluvainen 2005). For example, Maltamo et al. (2018) studied homogeneous *Eucalyptus urograndis* plantations and observed only marginal differences in the predictive performances between NN and PPM. The advantage of the parametric approaches is that they can be applied with a smaller training dataset. The parametric approaches can also extrapolate outside training data which is not the case with NN. We observed that GLM outperformed NN in the modeling of DBH distributions for deciduous-dominated plots, which may be explained by the lack of comprehensive training data for the NN model.

Because PPM and GLM produce relative DBH distributions, they also require a total stem frequency prediction from an additional model. In contrast, NN is based on the tree lists fetched from the nearest neighbors and can be used to predict DBH frequency distributions in a single step. This also means that several forest attributes, consistent with the DBH distribution, can be predicted based on the nearest neighbors.

Wall-to-wall maps of predicted forest attributes, based on NFI and remotely sensed data, are publicly available, for example, in Sweden (Nilsson et al. 2017), Finland (Kangas et al. 2018), and Norway (Astrup et al. 2019). However, wall-to-wall forest attribute maps may include systematic errors which can propagate in the aggregation of model-based predictions (e.g. grid cells) for large-area estimates. In MA estimation, systematic errors associated with the predicted forest attribute maps are mitigated (e.g., McRoberts et al. 2020, Næsset et al. 2020, Breidenbach et al. 2020b). This is a crucial step as can be seen from the non-zero correction factors utilized in the MA estimator (Table 4).

We estimated the study area DBH distribution using 2 cm and 6 cm bin widths. The RE values, and thus the efficiency gain achieved by the MA estimator, were larger when a bin width of 6 cm was used instead of 2 cm bin width. A DBH distribution with a bin width of 6 cm or even more may be appropriate for many applications focusing on large areas whereas a smaller bin width is typically required, for example, in forest management inventories. The efficiency gain achieved by the MA estimator compared with the direct estimator will also increase if the accuracies associated with the tree species map and the implicit forest/non-forest information could be improved further.

## 6 CONCLUSIONS

The following conclusions can be drawn from this study: i) GLM outperformed PPM in the prediction of DBH distributions whereas NN outperformed the parametric GLM and PPM approaches; ii) The use of NN-based predictions in the model-assisted estimation of the study area DBH distribution generally resulted in higher precisions of estimates compared with direct estimation; iii) The efficiency associated with the model-assisted estimator was larger when the DBH distribution was characterized using a bin width of 6 cm instead of a bin width of 2 cm.

## ACKNOWLEDGEMENTS

We would like to thank Marius Hauglin, Johannes Rahlf, and Johannes Schumacher for their support in the data preparation and processing. This study was supported by NIBIO (Norwegian Institute of Bioeconomy Research) and the PRECISION project (NFR# 11067).

## APPENDIX 1

### Variance estimator assuming simple random sampling

The variance estimator assuming simple random sampling (SRS) for the ratio $\hat{\mu}$ (Cochran 1977, p. 153–154; Mandallaz 2008, p. 63) is

$$\widehat{Var}_{SRS}(\hat{\mu}) = \frac{1}{n\left(\frac{\sum_{j\in S} I_j}{n}\right)^2}\frac{1}{n-1}\sum_{j\in S}\left(z_j\right)^2 \qquad (A1)$$

where $z_j = y_j - \hat{\mu}I_j$ in which $y_j$ is the observed stem frequency per hectare forest land at sample plot j belonging to sample S, $\hat{\mu}$ is the direct estimator for the mean stem frequency per forest land hectare (see Section 3.2 in the main document), and $n$ is the number of sample plots. The binary indicator variable $I_j$ is 1 for forested plots and 0 for non-forest plots, respectively.

The variance associated with the model-assisted estimator $\hat{\mu}_{MA}$ was estimated using the SRS estimator (Eq. A1) with

$$z_j = e_j - \bar{e} \qquad (A2)$$

where $e_j = y_j - \hat{y}_j$ is the residual associated with plot j and $\bar{e}$ is the mean of residuals.

Table A1 shows the characteristics associated with the estimation of the DBH distribution at the level of study area using the SRS variance estimator. Figure A1 shows the RE values associated with the model-assisted estimation of the DBH distribution using the SRS variance estimator. For more methodological details, please refer to Section 3.2 in the main paper.

**Table A1.** Characteristics of the direct and model-assisted estimates of the DBH distribution at the level of the study area. The variances were estimated using the simple random sampling (SRS) estimator. Model-assisted estimates were computed using the observed tree species and predicted tree species map (in parenthesis).

| DBH class (cm) * | Direct estimate $\hat{\mu}$ (stems per hectare) | Half 95% $CI_{SRS}$ $\hat{\mu}$, % | Half 95% $CI_{SRS}$ $\hat{\mu}_{MA}$, % | Correction factor $\hat{\mu}_C$, % | DBH class (cm) * | Direct estimate $\hat{\mu}$ (stems per hectare) | Half 95% $CI_{SRS}$ $\hat{\mu}$, % | Half 95% $CI_{SRS}$ $\hat{\mu}_{MA}$, % | Correction factor $\hat{\mu}_C$, % |
|---|---|---|---|---|---|---|---|---|---|
| 6 | 206.74 | 2.60 | 2.23 (2.34) | 3.16 (9.54) | 30 | 12.41 | 4.53 | 4.15 (4.21) | -2.46 (0.41) |
| 8 | 159.83 | 2.45 | 2.04 (2.13) | 0.81 (6.63) | 32 | 9.49 | 5.14 | 4.73 (4.74) | -4.54 (-1.17) |
| 10 | 121.24 | 2.42 | 1.99 (2.07) | -0.57 (4.32) | 34 | 7.38 | 5.66 | 5.37 (5.41) | -3.23 (0.71) |
| 12 | 92.04 | 2.46 | 2.09 (2.17) | -1.88 (2.22) | 36 | 5.52 | 6.34 | 6.12 (6.14) | -5.79 (-0.83) |
| 14 | 71.97 | 2.51 | 2.13 (2.24) | -1.03 (2.13) | 38 | 4.20 | 7.26 | 7.12 (7.14) | -0.11 (3.95) |
| 16 | 56.08 | 2.66 | 2.27 (2.36) | -2.41 (0.29) | 40 | 2.75 | 8.85 | 8.81 (8.86) | -11.93 (-7.79) |
| 18 | 44.70 | 2.87 | 2.47 (2.54) | -2.54 (-0.55) | 42 | 2.17 | 9.80 | 9.77 (9.80) | -0.16 (1.86) |
| 20 | 36.59 | 3.06 | 2.65 (2.71) | -1.42 (0.09) | 44 | 1.66 | 11.27 | 11.29 (11.25) | 4.73 (7.11) |
| 22 | 29.95 | 3.26 | 2.84 (2.91) | -2.20 (-0.50) | 46 | 1.04 | 13.56 | 13.82 (13.84) | -7.67 (-4.71) |
| 24 | 24.70 | 3.42 | 3.03 (3.09) | -0.42 (1.35) | 48 | 0.80 | 15.29 | 15.52 (15.51) | 0.65 (5.11) |
| 26 | 19.38 | 3.90 | 3.48 (3.54) | -4.16 (-1.59) | 50 | 0.60 | 19.15 | 19.42 (19.36) | 5.16 (8.12) |
| 28 | 15.78 | 4.14 | 3.72 (3.75) | -1.18 (1.16) | All | 928.23 | 1.60 | 1.12 (1.24) | 1.40 (4.23) |

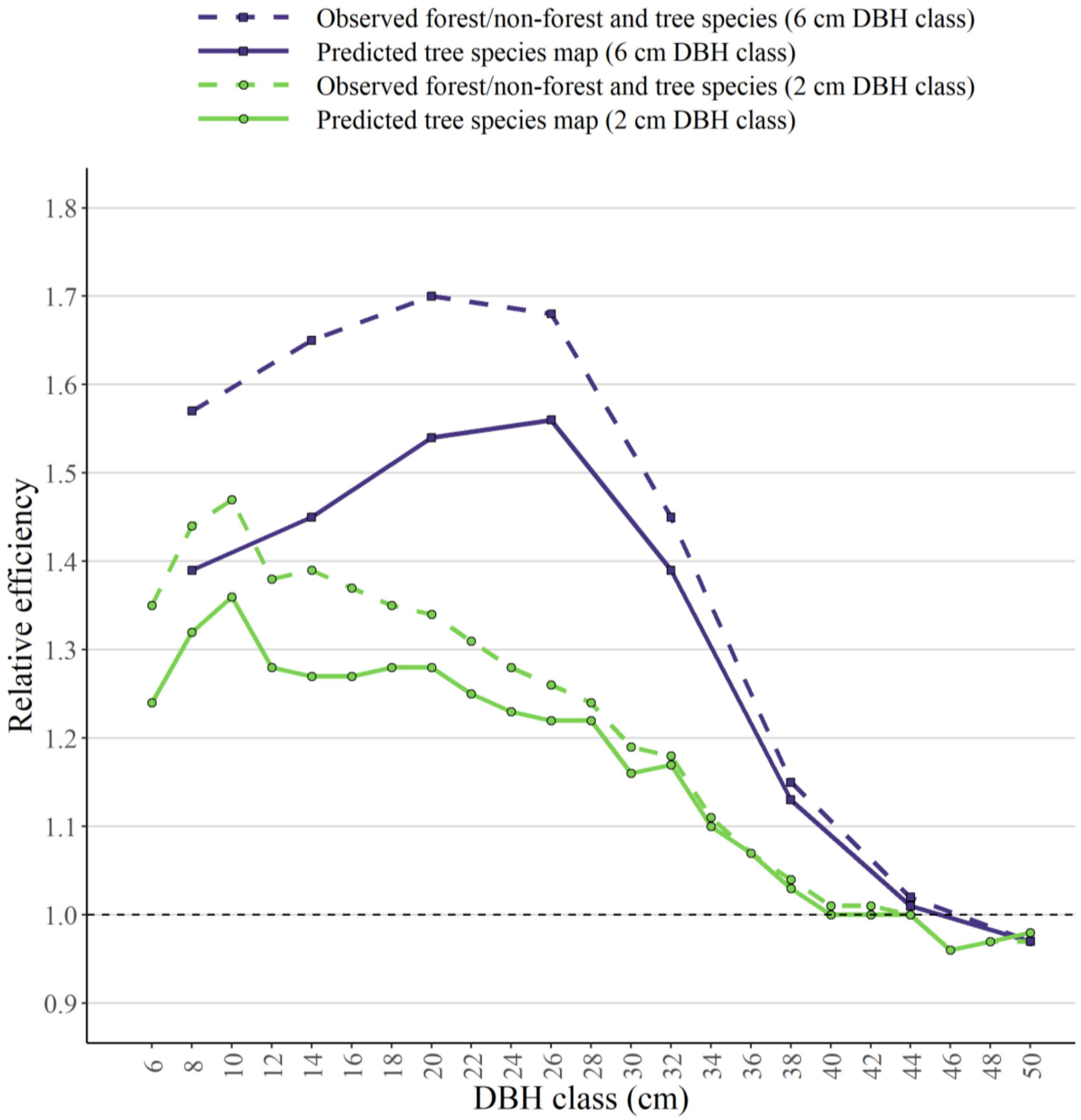

**Figure A1.** Relative efficiency (RE) of the model-assisted estimator by DBH classes. The variances were estimated assuming simple random sampling.

## APPENDIX 2

**Table A2.1.** Predictor variables of DBH distribution models by dominant tree species and modeling approaches. PPM – parameter prediction method using linear mixed-effects models, GLM – parameter prediction using generalized linear models, NN – Nearest neighbor approach

| Dominant tree species | NN | PPM/GLM Weibull scale | PPM/GLM Weibull shape |
|---|---|---|---|
| Norway spruce | hmean$_{first}$, h75$_{all}$, h95$_{all}$, proph$_{first}$, hskew$_{last}$ | proph$_{first}$, hskew$_{first}$, h95$_{all}$ | proph$_{first}$, hskew$_{first}$, hcv$_{first}$ |
| Scots pine | h25$_{first}$, h75$_{all}$, h95$_{first}$, d2$_{all}$, d6$_{first}$, | d4$_{all}$, h90$_{all}$, h50$_{last}$ | hmean$_{all}$, hvar$_{all}$, d6$_{first}$ |
| Deciduous species | h95$_{all}$, d0$_{first}$, d2$_{all}$, d4$_{first}$, hskew$_{first}$ | igratio$_{all}$, h50$_{all}$, hvar$_{first}$ | hvar$_{first}$, d8$_{first}$, h75$_{first}$ |

**Table A2.2.** Parameter estimates and their standard errors/confidence intervals (SE/CI) associated with the linear mixed-effects models (PPM) for Weibull scale and shape parameters. Please refer to Table 2 for the abbreviations of predictor variables. L – lower bound, U – upper bound

| Spruce-dominated | | | | | Pine-dominated | | | | | Deciduous-dominated | | | | |
|---|---|---|---|---|---|---|---|---|---|---|---|---|---|---|
| | scale | | shape | | | scale | | shape | | | scale | | shape | |
| | Est. | SE | Est. | SE | | Est. | SE | Est. | SE | | Est. | SE | Est. | SE |
| Intercept | 18.30 | 1.72 | 2.12 | 0.74 | Intercept | 3.53 | 1.05 | 1.33 | 0.17 | Intercept | 5.74 | 0.91 | 1.94 | 0.16 |
| $\text{proph}_{\text{first}}$ | -22.94 | 2.23 | -2.13 | 0.66 | $\text{d4}_{\text{all}}$ | 6.12 | 2.52 | - | - | $\text{igratio}_{\text{all}}$ | 1.21 | 1.21 | - | - |
| $\text{hskew}_{\text{first}}$ | -4.16 | 0.52 | -1.65 | 0.11 | $\text{h90}_{\text{all}}$ | 0.96 | 0.08 | - | - | $\text{h50}_{\text{all}}$ | 0.24 | 0.24 | - | - |
| $\text{h95}_{\text{all}}$ | 0.95 | 0.05 | - | - | $\text{h50}_{\text{last}}$ | -0.60 | 0.13 | - | - | $\text{hvar}_{\text{first}}$ | 0.20 | 0.20 | 0.26 | 0.01 |
| $\text{hcv}_{\text{first}}$ | - | - | 2.04 | 0.40 | $\text{hmean}_{\text{all}}$ | - | - | -0.30 | 0.05 | $\text{d8}_{\text{first}}$ | - | - | 3.69 | 1.10 |
| - | - | - | - | - | $\text{hvar}_{\text{all}}$ | - | - | 0.06 | 0.01 | $\text{h75}_{\text{first}}$ | - | - | -0.09 | 0.026 |
| - | - | - | - | - | $\text{d6}_{\text{first}}$ | - | - | 3.34 | 0.77 | - | - | - | - | - |
| | Est. | 95% CI [L, U] | Est. | 95% CI [L, U] | | Est. | 95% CI [L, U] | Est. | 95% CI [L, U] | | Est. | 95% CI [L, U] | Est. | 95% CI [L, U] |
| $\sigma_b$ * | 1.13 | [0.73, 1.74] | 0.08 | [0.02, 0.42] | $\sigma_b$ * | 1.86 | [1.27, 2.74] | 0.36 | [0.21, 0.60] | $\sigma_b$ * | 1.46 | [0.70, 3.06] | 0.17 | [0.02, 1.53] |
| $\sigma$ | 4.82 | [4.59, 5.06] | 0.89 | [0.85, 0.93] | $\sigma$ | 6.08 | [5.76, 6.41] | 1.43 | [1.36, 1.51] | $\sigma$ | 4.55 | [4.08, 5.06] | 0.98 | [0.88, 1.09] |

* Note: ALS project-area random effect

**Table A2.3.** Parameter estimates and their standard errors/confidence intervals (SE/CI) associated with the generalized linear models (GLM) for Weibull scale and shape parameters. Note that the log link function was used in the modeling of the Weibull parameters. Please refer to Table 2 for the abbreviations of predictor variables. L – lower bound, U – upper bound

| Spruce-dominated | | | | | Pine-dominated | | | | | Deciduous-dominated | | | | |
|---|---|---|---|---|---|---|---|---|---|---|---|---|---|---|
| | scale | | shape | | | scale | | shape | | | scale | | shape | |
| | Est. | SE | Est. | SE | | Est. | SE | Est. | SE | | Est. | SE | Est. | SE |
| Intercept | 2.60 | 0.03 | 1.98 | 0.16 | Intercept | 2.17 | 0.02 | 0.32 | 0.17 | Intercept | 1.91 | 0.02 | 0.51 | 0.02 |
| proph$_{first}$ | -1.01 | 0.03 | -1.82 | 0.13 | d4$_{all}$ | 0.12 | 0.04 | - | - | igratio$_{all}$ | 0.05 | 0.01 | - | - |
| hskew$_{first}$ | -0.16 | 0.01 | -0.56 | 0.02 | h90$_{all}$ | 0.05 | <0.01 | - | - | h50$_{all}$ | 0.02 | <0.01 | - | - |
| h95$_{all}$ | 0.06 | <0.01 | - | - | h50$_{last}$ | -0.03 | <0.01 | - | - | hvar$_{first}$ | 0.02 | <0.01 | <0.01 | <0.01 |
| hcv$_{first}$ | - | - | 0.04 | 0.09 | hmean$_{all}$ | - | - | -0.06 | <0.01 | d8$_{first}$ | - | - | 1.46 | 0.12 |
| - | - | - | - | - | hvar$_{all}$ | - | - | 0.01 | <0.01 | h75$_{first}$ | - | - | -0.02 | <0.01 |
| - | - | - | - | - | d6$_{first}$ | - | - | 1.26 | 0.08 | - | - | - | - | - |
| | Est. | 95% CI [L, U] | Est. | 95% CI [L, U] | | Est. | 95% CI [L, U] | Est. | 95% CI [L, U] | | Est. | 95% CI [L, U] | Est. | 95% CI [L, U] |
| $\sigma_b$ | 0.19 | [0.18, 0.20] | 0.20 | [0.19, 0.22] | $\sigma_b$ | 0.22 | [0.21, 0.23] | 0.31 | [0.30, 0.33] | $\sigma_b$ | 0.23 | [0.21, 0.26] | * | * |

* Note: plot-level random effect was not used for the shape parameter due to numerical instability